\documentclass[pre,aps,twocolumn,amsmath,amssymb,showpacs,showkeys,unsortedaddress]{revtex4-1}
\usepackage{graphicx}
\usepackage{psfrag}
\usepackage{epsfig}
\usepackage{dcolumn}
\usepackage{bm}

\newcommand{\eq}[1]{Eq.~(\ref{#1})}    
\newcommand{\fg}[1]{Fig.~\ref{#1}}     

\begin{document}
\title{Opinion formation in the kinetic exchange models: Spontaneous symmetry breaking transition}

\author{Mehdi Lallouache}
  \email{mehdi.lallouache@ens-cachan.fr}
  \affiliation{D\'epartement de Physique, \'Ecole Normale Sup\'erieure de Cachan, 94230 Cachan, France}
  \affiliation{Chaire de Finance Quantitative, Laboratoire de Math\'ematiques Appliqu\'ees aux Syst\`emes, \'Ecole Centrale Paris, 92290 Ch\^atenay-Malabry, France}
  
  \author{Anindya S. Chakrabarti}
  \email{aschakrabarti@gmail.com}
  \affiliation{Department of Economics, Boston University, 270 Bay State Road, Boston, Massachusetts 02215, USA}

\author{Anirban Chakraborti}
  \email{anirban.chakraborti@ecp.fr}
   \affiliation{Chaire de Finance Quantitative, Laboratoire de Math\'ematiques Appliqu\'ees aux Syst\`emes, \'Ecole Centrale Paris, 92290 Ch\^atenay-Malabry, France}
   
\author{Bikas K. Chakrabarti}
 \email{bikask.chakrabarti@saha.ac.in}
 \affiliation{Centre for Applied Mathematics and Computational Science,
Saha Institute of Nuclear Physics,
1/AF Bidhannagar, Kolkata 700 064, India}
\affiliation{Economic Research Unit, Indian Statistical Institute, 203 B. T. Road, Kolkata 700 018, India}

\date{\today}

\begin{abstract}
We propose a minimal multi-agent model for the collective dynamics of opinion formation
in the society, by modifying kinetic exchange dynamics studied in the context of
income, money or wealth distributions in a society. This model has an intriguing spontaneous symmetry
breaking transition to polarized opinion state starting from non-polarized opinion state. In order to analyze the model, we introduce an iterative map version of the model, which has very similar statistical characteristics. An approximate theoretical analysis of the numerical results are also given, based on the iterative map version.
\end{abstract}


\maketitle

\section{Introduction}
\label{intro}
Recently physicists have been studying social phenomena and
dynamics leading to the growth of the interdisciplinary field of ``Sociophysics'' \cite{CCCbook2006}. One of the problems is of ``opinion formation'', which is a collective dynamical phenomenon,
and as such is closely related to the problems of competing cultures or languages \cite{Castellano2009,StaufferWeisbuchGalam2006}. 
It deals with a ``measurable'' response of the society to e.g., political issues, acceptances of
innovations, etc. Numerous models of competing options have been introduced
to study this phenomenon, e.g., the ``voter'' model (which has a binary opinion variable with the
opinion alignment proceeding by a random choice of neighbors) \cite{Holley1975}, or the Sznajd-Weron discrete opinion formation model (where
more than just a pair of spins is associated with the decision
making procedure) \cite{Sznajd-Weron2000}. 
There have been other studies of systems with more than just two possible opinions \cite{Vazquez2003}, or where the opinion of individuals is represented
by a ``continuous'' variable \cite{Hegselman2002,Deffuant2000,Fortunato2005} using real numbers. Also,
since opinion formation in a human society is mediated by social
interactions between individuals,  such social dynamics has been considered to take
place on a network of relationships (see \cite{Castellano2009} for recent review on such models). 

A two body exchange dynamics has already been developed in the context of modelling
income, money or wealth distributions in a society \cite{Yakovenko2009,Arnab2007,Patriarca2010,Chakraborti2010a,Redner2010}.
The general aim was to study a many-agent statistical model
of closed economy (analogous to the kinetic theory model of ideal gases) \cite{Chakraborti2000a},
where $N$ agents exchange a quantity $x$, that
may be defined as wealth.
The states of agents are characterized by the wealth
$\{x_i\},~i=1,2,\dots,N$, such that $x_i >0, \quad \forall i$ and the total wealth $W=\sum_{i} x_i$ is conserved.
The question of interest is: ``What is the equilibrium distribution of wealth $f(x)$, such that $f(x) dx $ is the probability that in the steady state of the system, a randomly chosen agent will be found to have wealth between $x$ and $x + dx$?''
The evolution of the system is carried out according to a prescription,
which defines the trading rule between agents. The agents 
interact with each other through a pair-wise interaction characterized
by a ``saving'' parameter $\lambda$, with $0 \le \lambda \le 1$. 
The dynamics of the model (CC)
is as follows
\cite{Chakraborti2000a}:
\begin{eqnarray}
  x_i' &=& \lambda x_i + \epsilon (1-\lambda) (x_i + x_j) \, ,
  \nonumber \\
  x_j' &=& \lambda x_j + (1-\epsilon) (1-\lambda) (x_i + x_j) \, ,
  \label{sp1}
\end{eqnarray}
where $\epsilon$ $(0 \le \epsilon \le 1)$ is a stochastic variable, changing with time.
It can be noticed that in this way, the quantity $x$ is conserved
during the single transactions: $x_i'+x_j' = x_i + x_j$,
where $x_i'$ and $x_j'$ are the agent wealths
after the transaction has taken place.
In general, the functional form for steady state distribution $f(x)$ is seen to be close to the $\Gamma$-distribution \cite{Anirban2008,Patriarca2004a}.
As a further generalization, the agents could be assigned different saving propensities and the steady state distribution $f(x)$ show Pareto-like power-law behavior asymptotically \cite{Chatterjee2004a, Chakraborti2009}.


Earlier, Toscani \cite{Toscani2006} had introduced and discussed kinetic models of (continuous) opinion formation involving both exchange of opinion between individual agents and diffusion of information. Based on this model, During et al \cite{During2009} proposed another mathematical model for opinion formation in a society that is built of two groups, one group of ‘ordinary’ people and one group of ‘strong opinion leaders’. Starting from microscopic interactions among individuals, they arrived at a macroscopic description of the opinion formation process. They discussed the steady states of the system, and extended it to incorporate emergence and decline of opinion leaders. 

Here, we report the studies of a minimal model for the collective dynamics of opinion formation
in the society, based on kinetic exchanges.

\section{Model for opinion formation and results}
\label{model}

\subsection{Homogeneous multi-agent model}
\label{multi}

Following the CC model described in the earlier section, we present a minimal model \cite{Mehdi2010} for the collective dynamics of opinion $ O_i(t) $ of the $ i $-th person in the society, consisting of $ N $ ($ N \longrightarrow \infty$) persons. We assume that any particular person can discuss (interact) only with one other person each time (time increases discretely by unity after each such discussion). A two-person ``discussion'' is viewed here as a simple two-body \textit{scattering process} in physics. Persons in the society may bump onto each other randomly and exchange opinions through such random two-person discussions. In general, a person $ i $ could have any opinion $ O_i $ between two extreme polarities denoted by  +1 and -1.
In any discussion at time $ t+1 $, a person \textit{retains} a fraction of his/her older opinion $O_i(t)$, determined by his/her ``conviction'', parameterized by $\lambda_i$. This parameter value is characteristic of a person and does not change with time $ t $. Additionally, the person $ i $ is ``influenced'' stochastically by the other person $ j $ during the discussion having the ``influence'' parameter equal to his/her conviction parameter $\lambda_j$.
We further assume for simplicity that all agents are \textit{homogeneous} -- have the \textit{same} conviction parameter $\lambda$. Mathematically the dynamics may be represented by
\begin{align}
  O_i(t+1) &= \lambda (O_i(t) + \epsilon_t O_j(t)) \, ,
  \nonumber \\
  O_j(t+1) &= \lambda (O_j(t) + \epsilon'_t O_i(t)) \, ,
  \label{op2}
\end{align}
where the opinion $-1 \le O_i(t) \le 1$ for all agents $i$ and time $t$, the conviction parameter $0 \le \lambda \le 1$ is \textit{quenched} (does not change with time), and the stochastic parameters $\epsilon_t$ and $\epsilon'_t$ are \textit{annealed} variables (change with time)-- uncorrelated random numbers uniformly distributed between zero and unity. Note that the equations are linear, but non-linearity is introduced in this model by imposing that $-1\le O_i(t)\le 1$ for all agents $i$ and times $t$.

\begin{figure}
\begin{center}
    \psfrag{ylabel}{$\bar{O}$}
    \psfrag{xlabel}{$\lambda$}
    \psfrag{insetylabel}{$(\Delta O)^2$}
    \psfrag{insetxlabel}{$\lambda$}
	\includegraphics[width=0.85\linewidth]
        {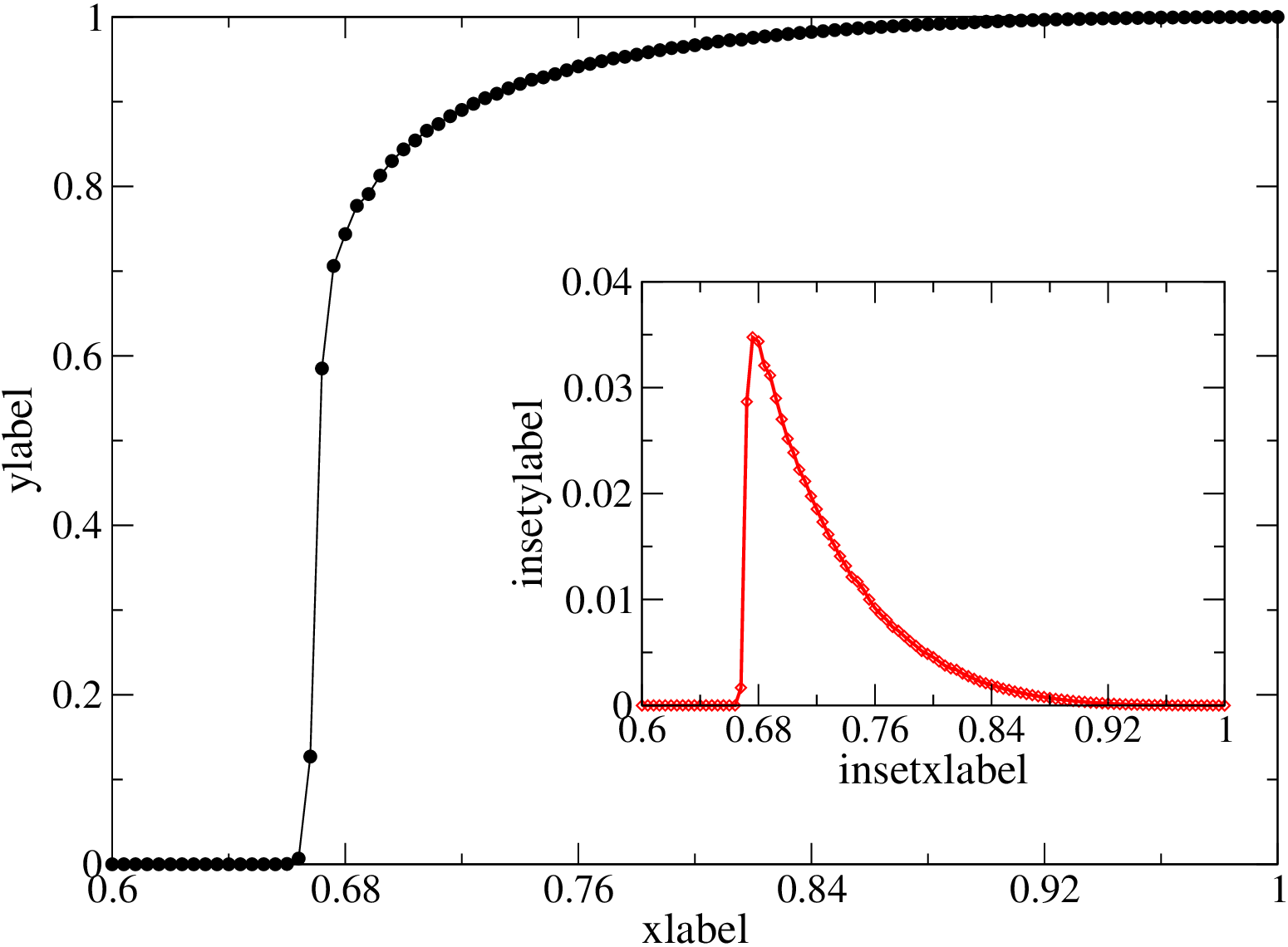}
    \end{center}
\caption{Numerical results for the variation of the average opinion $\bar{O}(t)$ for large $t$ (steady state value of $\bar{O}$) against $\lambda$, following dynamics of \eq{op2}.
(Inset) Numerical results for the variation of the variance $(\Delta O)^2 \equiv \overline{(O-\bar{O})^2}$ against $\lambda$, following dynamics of \eq{op2}.
}
\label{fig:phase_multi}
\end{figure}

The question we are interested is that if such social dynamics continually take place, can any consensus be reached or polarity evolve after a long time?
Mathematically, we are interested in the steady state distribution of $O$ and other statistical properties.
It is noteworthy that unlike in the market models, here we have no conservation of opinion. Rather, the steady state of value of $\bar{O}(t)= (1/N) |\Sigma_i O_i(t)|$ represents the order of the average opinion in the society after a long time $t$. We study the relaxation dynamics in the society: the relaxation and fluctuation of $\bar{O}$, the steady state value of $\bar{O}(t)$ for $t>\tau$, the relaxation time.

Remarkably, we find there is appearance of ``polarity'' or consensus, starting from initial random disorder (where $O_i$'s are uniformly distributed with positive and negative values). In the language of physics, there is a ``spontaneous symmetry breaking'' transition in the system: starting from $\bar{O}(0)=0$
the system evolves either to the
``para'' state with $\bar{O} \equiv \bar{O} (t > \tau)=0$ (where all individual agents 
have the opinion $0$) for $\lambda \le 2/3$, or (\textit{continuously}) to the ``symmetry broken'' state $\bar{O} \equiv \bar{O} (t > \tau) \ne 0$ (where \textit{all} individuals have either positive or negative opinions) for $\lambda \ge 2/3$ (see \fg{fig:phase_multi}) for times $t > \tau$.  We note, however, that the fluctuation in $\bar{O}$ does not diverge, and shows a cusp near $\lambda_c$ (see inset of \fg{fig:phase_multi}).  We also study the relaxation behavior of $\bar{O}(t)$ and the critical divergence of the relaxation time $\tau$ near $\lambda=\lambda_c=2/3$ (see subsection \ref{analyses}, \fg{fig:relaxmap}).

\subsection{Random multiplier map}
\label{map}

\begin{figure}
\begin{center}
    \psfrag{ylabel}{$\bar{O}$}
    \psfrag{xlabel}{$\lambda$}
    \psfrag{insetylabel}{$(\Delta O)^2$}
    \psfrag{insetxlabel}{$\lambda$}
	\includegraphics[width=0.85\linewidth]
        {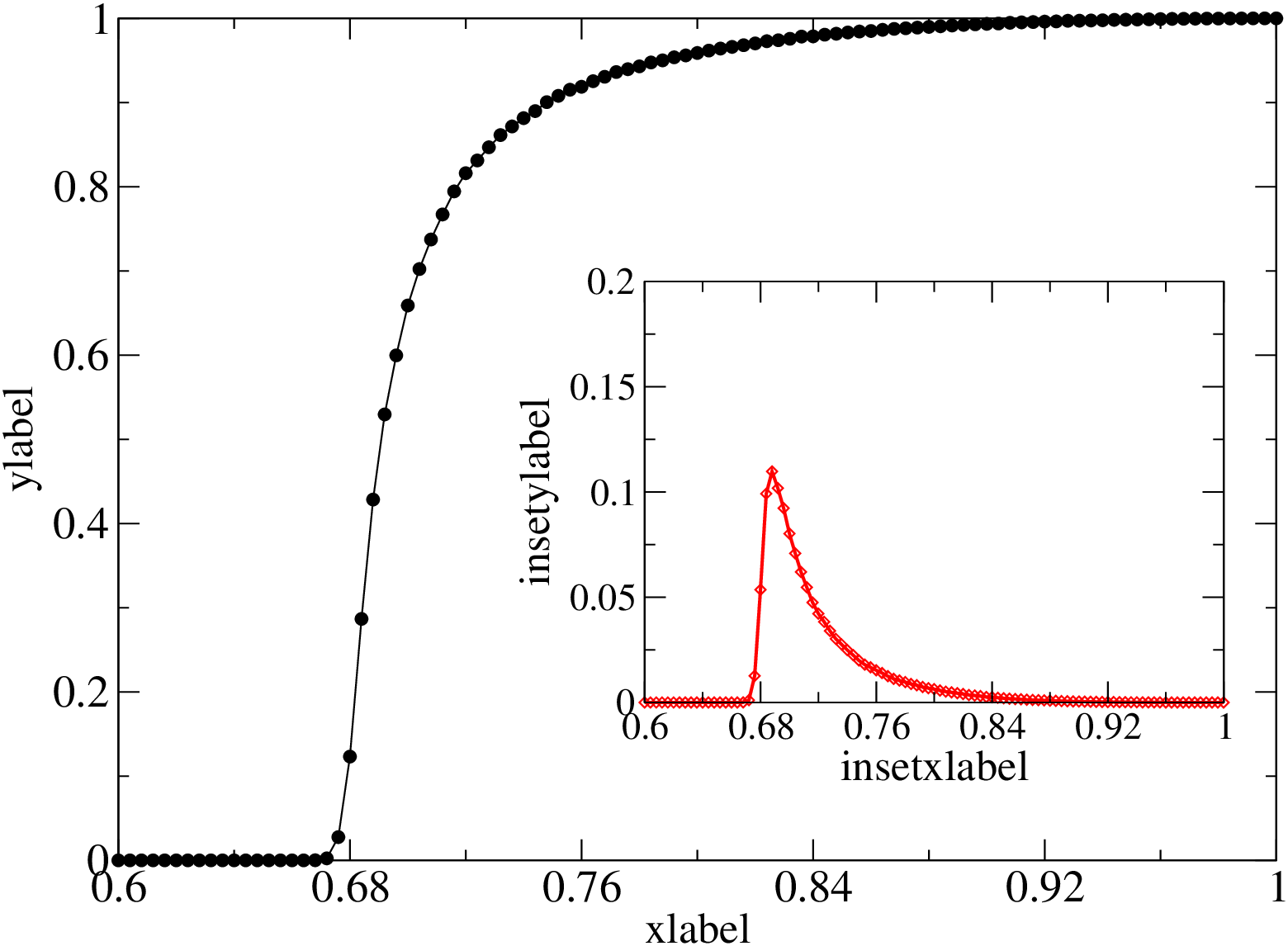}
    \end{center}
\caption{Numerical results for the variation of the average opinion $\bar{O}(t)$ for large $t$ (steady state value of $\bar{O}$) against $\lambda$, following dynamics of \eq{op3}.
(Inset) Numerical results for the variation of the variance $(\Delta O)^2 \equiv \overline{(O-\bar{O})^2}$ against $\lambda$, following dynamics of \eq{op3}.
}
\label{fig:phase_map}
\end{figure}

The basic nature of transition produced by \eq{op2}, can perhaps be reproduced by the following simple iterative map
\begin{equation} 
O(t+1)=\lambda(1+\epsilon_t)O(t)
\label{op3}
\end{equation}
with the restriction that $O(t) \le 1$, which is ensured by assuming that if $O(t)\ge 1$, $O(t)$ is set equal to 1. As usual, $\epsilon_t$ is a stochastic variable ranging between 0 and 1 (assumed to be uniformly distributed in our case). In a mean-field approach, the above equation reduces effectively to a multiplier map like $O(t+1)=\lambda (1+ \langle \epsilon \rangle) O(t)$, where $\langle \epsilon \rangle=1/2$. Clearly for $\lambda\le 2/3$, $O(t)$ converges to zero. The initial value $ O(0) $, is assigned either a positive or negative value. If it starts from a positive (negative) value, $ O(t) $ remains positive (negative). 
We note that there are subtle differences in the dynamics of \eq{op2} and \eq{op3}. Apart from the absence of ``spontaneous symmetry breaking'' of the multi-agent model (from $\pm O_i(0)$ values to all positive or all negative transition beyond $\lambda_c$), the nature of the phase transition (singularity) in the iterative map is also slightly different. The critical value $\lambda_c=\exp{\lbrace -(2 \ln 2 -1) \rbrace} \approx 0.6796$ has an analytical derivation \cite{marsili}, but for most numerical studies done here, we take $\lambda_c=0.68$. The time variation of the average opinion $\bar{O}(t)=(1/N)\Sigma_i |O_i(t)|$, where $i$ refers to different initial realizations and $ N $ refers to the total of all such realizations, and its fluctuations are studied numerically (see \fg{fig:phase_map}). 
We study the relaxation behavior of $\bar{O}(t)$ and the critical divergence of the relaxation time $\tau$ near $\lambda=\lambda_c=0.68$ (see subsection \ref{analyses}, \fg{fig:relaxmap}). We note again that the fluctuation in $\bar{O}$ does not diverge, and shows a cusp near $\lambda_c=0.68$ (see inset of \fg{fig:phase_map}).
We also note that the steady state fluctuation $\Delta O$ near $\lambda_c$, is generally much higher in magnitude for the map case. 

\subsection{Results and analyses}
\label{analyses}

For both the multi-agent model and the iterative map, we study the variation (with $\lambda$) of the fraction $p$ of the agents having $O_i=\pm 1$ at any time $t$ in the steady state ($t>\tau$). This parameter $p$ gives the average ``condensation'' fraction (of people in the society having extreme opinions $|O_i|=1$) in the steady state.
The growth of $ p $,  as shown in \fg{fig:phase_map-multi}, is seen to be similar to that of $\bar{O}$. The inset shows that the growth behavior for $ p $ above (respective) $\lambda_c$, for both the multi-agent model and map, are identical.

We studied the relaxation behavior of $\bar{O}$ and $ p $, for both the multi-agent model and map. In each case, the relaxation time is estimated numerically from the time value at which $\bar{O}$ or $p$ first touches the steady state value within a pre-assigned error limit. We find diverging growth of relaxation time $\tau$ (for both $\bar{O}$ and $p$) near $\lambda=\lambda_c$ (see \fg{fig:relaxmap}).
The values of exponent $ z $ for the divergence in $\tau \sim |\lambda-\lambda_c|^{-z}$ have been estimated numerically for both the multi-agent model and the map (for both $\lambda>\lambda_c$ and $\lambda<\lambda_c$, wherever accurate data were obtained).
For the multi agent model, the fitting values for exponent $ z $ corresponding to $\bar{O}$ and $p$, respectively, are $z \approx 1.0 \pm 0.1$ and $z \approx 0.7 \pm 0.1$. 
For the map case, the fitting values for exponent $ z $ corresponding to both $\bar{O}$ and $p$, turn out to be the same: $z \approx 1.5 \pm 0.1$.

For the iterative map \eq{op3}, we study carefully the time evolution of the condensation fraction $p$ of $|O|=1$ in different realizations at different values of $\lambda$. The variation of the steady state value $p$ against $\lambda$ is shown in \fg{fig:phase_map}. It may be noted that while the steady state value of $\bar{O}$ starts to grow from $\lambda \approx 2/3$ (see \fg{fig:phase_map}) , the steady state value of $p$ starts growing at $\lambda \approx  0.68$ (see \fg{fig:phase_map-multi}). Numerical results for the growth of the relaxation time $\tau$ for both $\bar{O}$ and $p$, against $\lambda$ are shown in \fg{fig:relaxmap}. Both diverge at $\lambda \approx 0.68$. This clearly indicates that $p$, rather than $\bar{O}$, is the order parameter for the transition. 

\begin{figure}
\begin{center}
    \psfrag{ylabel}{$p$}
    \psfrag{xlabel}{$\lambda$}
    \psfrag{insetylabel}{$p$}
    \psfrag{insetxlabel}{$\lambda-\lambda_c$}
	\includegraphics[width=0.85\linewidth]
        {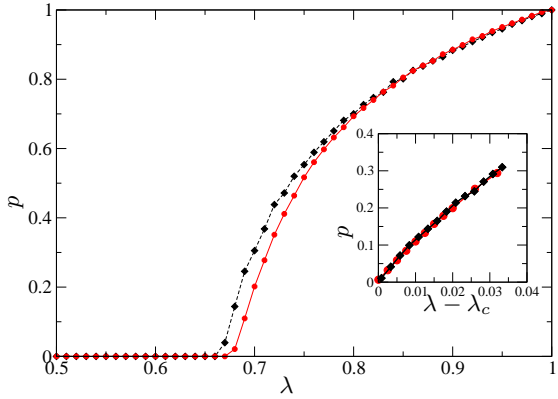}
    \end{center}
\caption{Numerical results for the variation of the average condensate fraction $p(t)$ for large $t$ (steady state value of $p$) against $\lambda$, following dynamics of \eq{op2} in black diamonds, and dynamics of \eq{op3} in red circles.
(Inset) Numerical results for the growth of $ p $, following dynamics of \eq{op2} in black diamonds, and dynamics of \eq{op3} in red circles, close to $\lambda_c$.
}
\label{fig:phase_map-multi}
\end{figure}


\begin{figure}
        \psfrag{ylabel}{$\tau$}
        \psfrag{xlabel}{$ ( \lambda-\lambda_c) $}
        \psfrag{insetylabel}{$ \tau $}
        \psfrag{insetxlabel}{$ | \lambda-\lambda_c| $}
	    \includegraphics[width=0.7\linewidth]
        {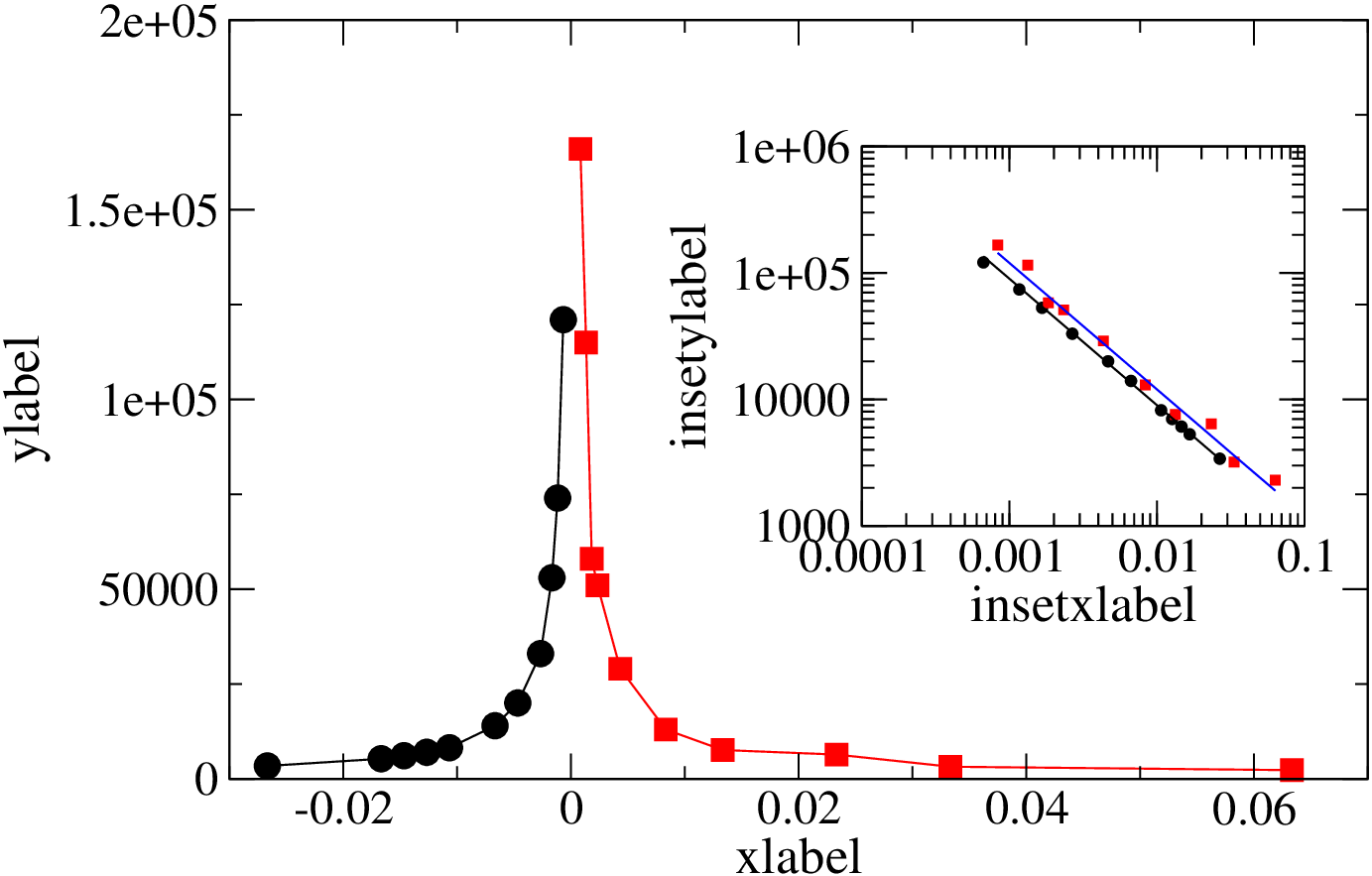}\\
        \vskip 0.45cm
        \includegraphics[width=0.7\linewidth]
        {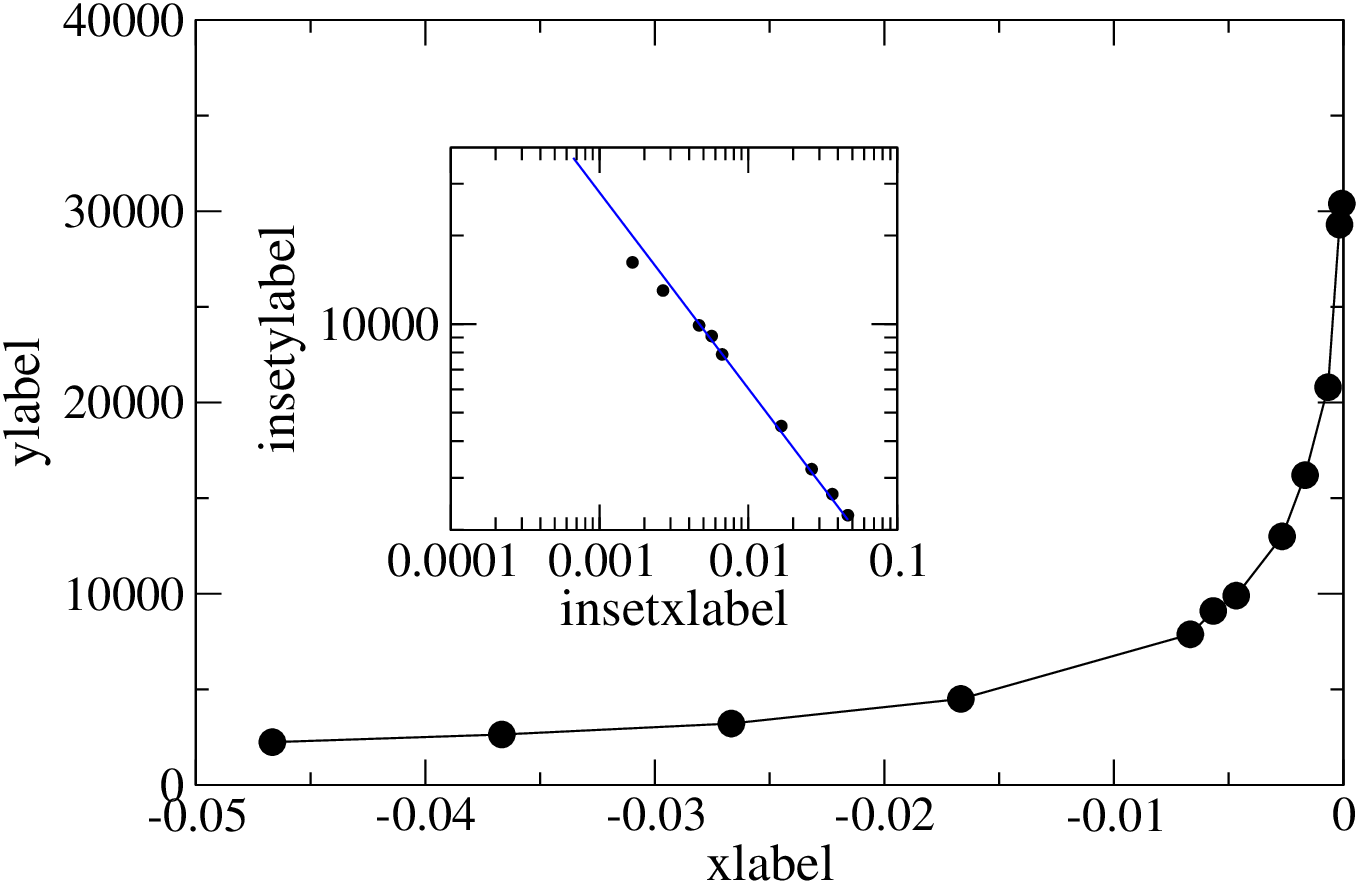}\\
        \vskip 0.45cm
        \includegraphics[width=0.7\linewidth]
        {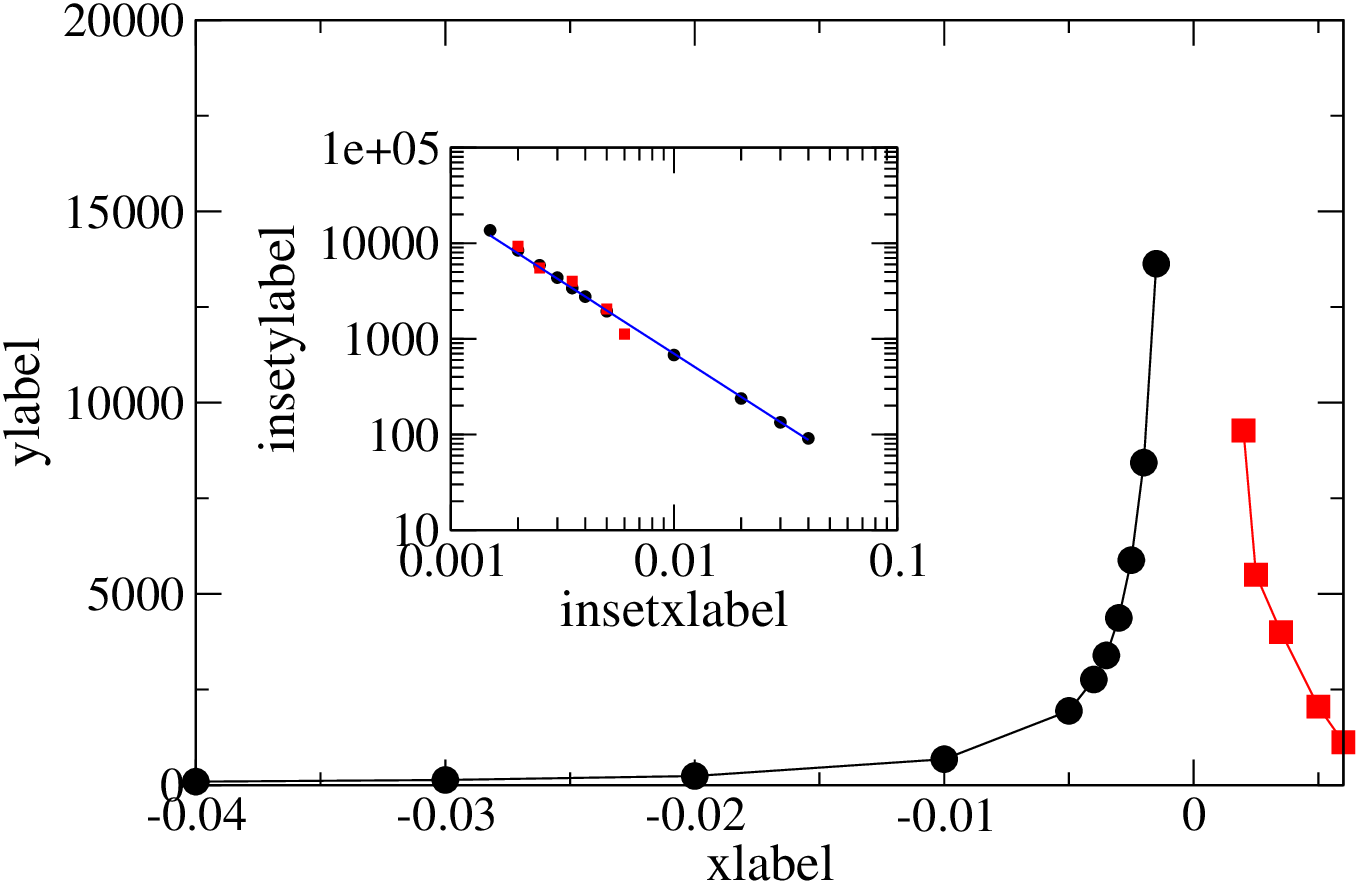}\\
        \vskip 0.45cm
        \includegraphics[width=0.7\linewidth]
        {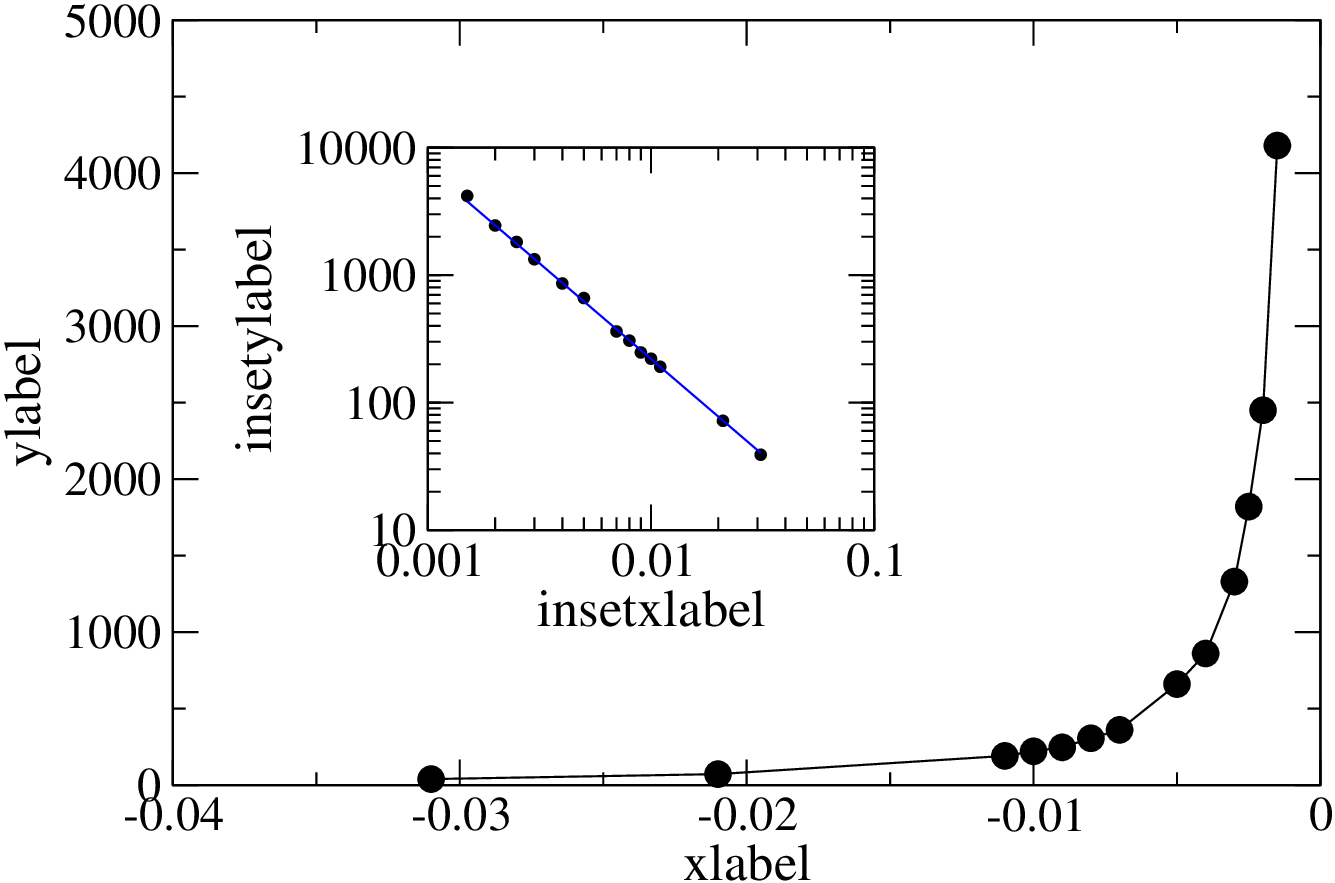}\\
\caption{Numerical results for relaxation time behaviors $ \tau $ versus $ \lambda- \lambda_c $, for (a) Multi-agent model with $ \bar{O} $ (b) Multi-agent model with $ p $ (c) Map with $ \bar{O} $ (d) Map with $ p $ . (Insets) Determination of exponent $ z $ from numerical fits of $\tau \sim | \lambda-\lambda_c|^{-z}$.}
\label{fig:relaxmap}
\end{figure}

\begin{figure}
\begin{center}
    \psfrag{ylabel}{$ P(|O|) $}
    \psfrag{xlabel}{$ |O| $}
    \psfrag{insetylabel}{$ P(|O|) $}
    \psfrag{insetxlabel}{$ |O| $}
	\includegraphics[width=0.75\linewidth]
        {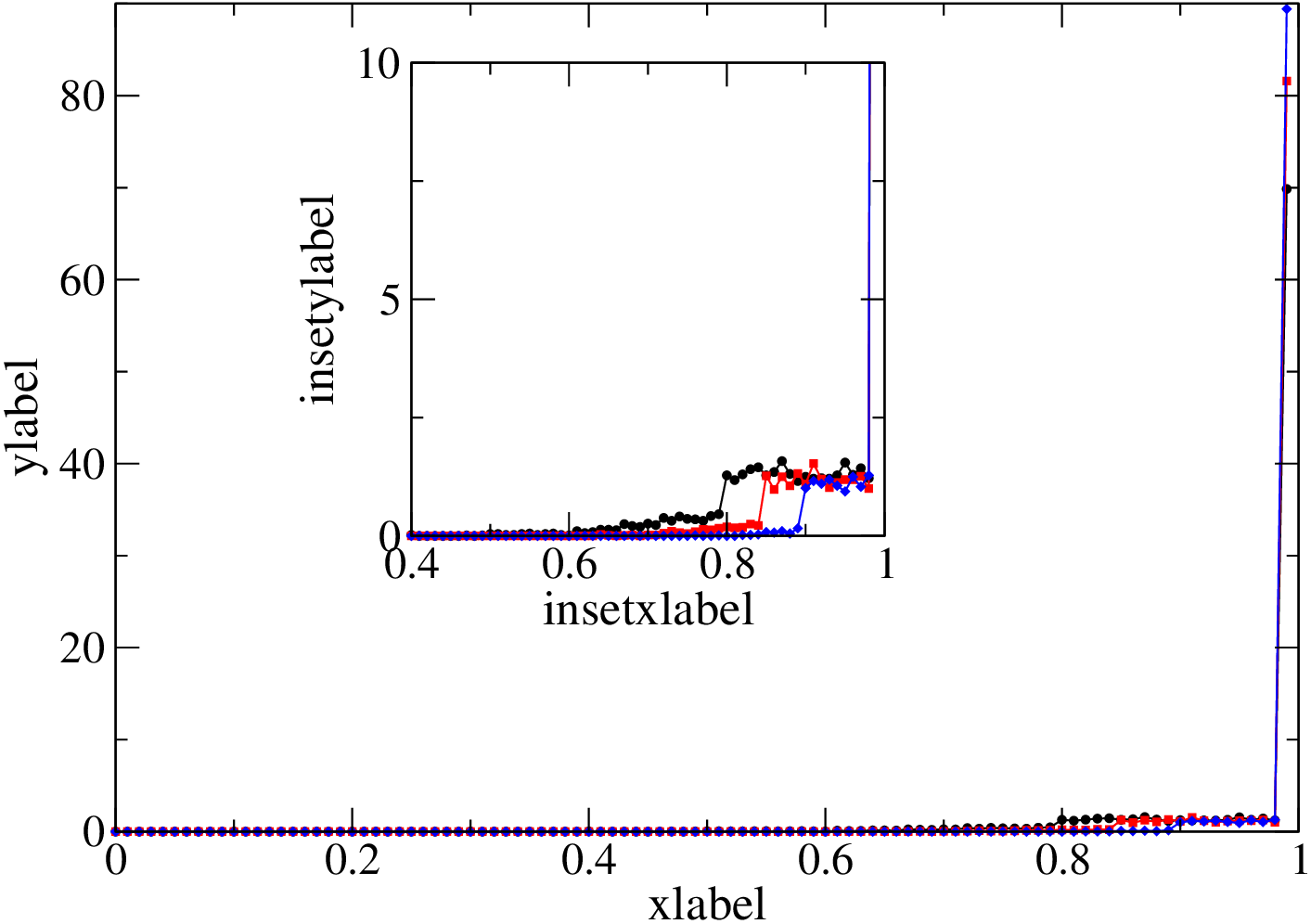}
    \end{center}
\caption{Numerical results for for the steady state distribution opinion, $ P(|O|) $,
for three values $\lambda=0.8,0.85,0.9$ showing bi-modal distributions in each case.
(Inset). The same steady state distribution $ P(|O|) $,
for three values $\lambda=0.8,0.85,0.9$, but close to $|O|=1$.
}
\label{fig:hist}
\end{figure}

An approximate analysis of the above transition for $\lambda$ closer to unity can be done for the iterative map \eq{op3} as follows. 
In \fg{fig:hist}, we give the numerical results for the steady state distribution opinion, $ P(|O|) $
for three different values of $\lambda$; we observe roughly a bi-modal nature of the distribution as $\lambda \rightarrow 1$: one mode is the uniform distribution within the range $ |O_{min}| < |O| < 1 $ (and $ |O_{min}| \approx \lambda $) and another a delta function at $ |O|=1 $. 
We therefore approximate the steady state distribution of opinion by assuming that opinion $O(t)$ is distributed uniformly starting from a minimum $O_{min}$ upto unity with (integrated) probability $(1-p)$, and a $\delta$-function at exactly unity with probability $p$. Then 
\begin{equation}
\bar{O}=(1-p)O_{av}+p. 1,
\label{av opi}
\end{equation}
where $O_{av}=(O_{min}+1)/2$.
We have assumed that the value $ O(t) $ stays in those two regions (from $\lambda$ to 1 and exactly at 1) 
with probability $(1-p)$ and $p$. Hence,
the corresponding equations are
\begin{equation} O(t+1)=\lambda(1+\epsilon)O(t) \hspace{1 cm} \mbox{with probability $1-p$} \nonumber ,
\end{equation}
and
\begin{equation} O(t+1)=1  \hspace{1 cm} \mbox{with probability $p$} \nonumber .
\end{equation}

Note that the first equation is realized only if
$\lambda(1+\epsilon)O(t)<1 $
or $\epsilon<\epsilon_{max}=\frac{1}{\lambda O_{av}}-1.$
This cut-off implies that
$p= \int_0^{\epsilon_{max}}d\epsilon = \frac{1}{\lambda O_{av}} -1$, since $\epsilon\sim$uni[0,1].
By substituting $O_{av}$ and $p$ in \eq{av opi}, we derive
the result that
\begin{eqnarray}
\bar{O} 
                 &=& \frac{5\lambda+2\lambda^2-\lambda^3-2}{2\lambda(1+\lambda)}
\label{maverage}
\end{eqnarray}
which is compared with the numerical simulations for $\lambda \rightarrow 1$ in \fg{fig:fit}. It is evident that the approximation holds well, only for $\lambda \rightarrow 1$.

\begin{figure}
\begin{center}
    \psfrag{ylabel}{$\bar{O}$}
    \psfrag{xlabel}{$\lambda$}
\includegraphics[width=0.85\linewidth]
	{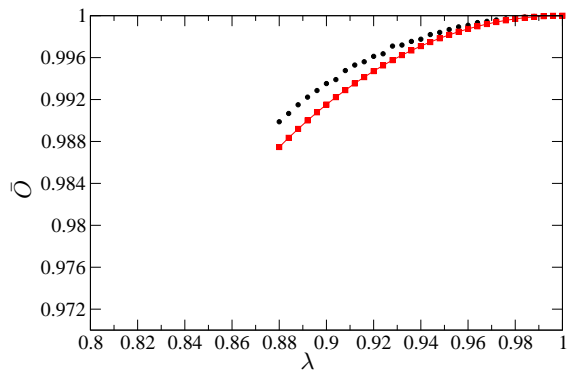}
\end{center}
\caption{Fit of the approximate theoretical calculation \eq{maverage} (in red squares) with the numerical simulations for $\lambda \rightarrow 1$, following dynamics of \eq{op3} (in black circles). }
\label{fig:fit}
\end{figure}

\section{Summary and Discussion}
\label{sec:summary}
In summary, we proposed a minimal model for the collective dynamics of opinion formation
in the society, by modifying kinetic exchange dynamics studied in the context of
markets. 
The multi-agent model (dynamics given by \eq{op2}) and its map version (dynamics given by \eq{op3}) have kinetic exchange like linear contributions from random two-person discussions or scattering processes, though the saturation of $ |O_i| \leq 1 $ induces non-linearity in the dynamics.
This model has an intriguing spontaneous symmetry
breaking transition to polarized opinion state starting from non-polarized opinion state.
Specifically, in the multi-agent model, we see that for $\lambda > \lambda_c = 2/3$, starting from random positive and negative $ O_i $ values (or for that matter any arbitrary state), at $ t=0 $, the system eventually evolves to a state at $ t> \tau $ where \textit{all} $ O_i $ are either positive or negative, with $ |\bar{O} | $ determined by the $ \lambda $ value! This is similar to the growth of spontaneous magnetization in Ising magnets (where starting from arbitrary up and down spin states, a preferred direction is chosen by fluctuation), with magnetization determined by the temperature below its transition value.
The appearance of spontaneous symmetry breaking in this simple kinetic opinion exchange model is truly remarkable. It
appears to be one of the simplest collective dynamical model of many-body dynamics showing non-trivial phase
transition behaviour. 
Indeed, it may be noted that for $\lambda \leq \lambda_c$, at $ t> \tau $, \textit{all} $ O_i $'s become identically zero (without any fluctuation), while for $\lambda > \lambda_c$, $ O_i $'s have fluctuations but the average has a steady state value depending on the value of $ \lambda $. As such the nature of the phase transition in this model is quite different and does not fit to the commonly studied three absorbing state models studied (see e.g., \cite{munoz}).
In order to understand the nature of the transition, we also studied a simple iterative map and derived approximate result for the order parameter variation under certain limits, which compares quite well with the numerical simulations. 
Specifically, we find that the fraction $ p $ of people with extreme opinion $ |O_i| = 1 $, and its fluctuations determine the nature of the phase transition in our model and locate the critical point accurately (from numerical studies).
With the two mode distribution (uniform and delta) of $ O $, valid close to $ \lambda \rightarrow 1 $ (see \fg{fig:hist}), we could develop an approximate analysis of the variation of the steady state mean opinion $ |\bar{O} | $ against $ \lambda$ as in \eq{maverage}.
In any case, further investigations are necessary for understanding this phase transition. Additional studies for the \textit{heterogeneous} conviction factors $ \lambda_i $'s, in influence of ``field terms'' that represent the external influence of media, etc. will be reported elsewhere \cite{Kaski2010}. Also, the study of this phase transition behavior for an extended model with separate``conviction'' and ``influence'' parameters in \eq{op2}, has recently been reported \cite{sen2010}.

\begin{acknowledgments}
The authors acknowledge F. Abergel, A. Chatterjee, A. Jedidi, K. Kaski, S.N. Majumdar, M. Marsili, N. Millot, M.A. Mu\~{n}oz, M. Patriarca and S. Pradhan for useful discussions, comments and correspondences.
\end{acknowledgments}

\end{document}